\documentclass[aps,prl,twocolumn,superscriptaddress,groupedaddress,preprintnumbers]{revtex4} 
\usepackage{graphicx}  
\usepackage{dcolumn}   
\usepackage{bm}        
\usepackage{amssymb}   
\usepackage{amsmath}
\usepackage{mathtools}
\usepackage{cases}
\usepackage{mathrsfs}
\usepackage{color}
\usepackage{enumitem}
\usepackage{ulem}

\newcommand{\mpl}{M_{\text{Pl}}}

\newcommand{\be}{\begin{equation}}
\newcommand{\ee}{\end{equation}}
\newcommand{\bea}{\begin{eqnarray}}
\newcommand{\eea}{\end{eqnarray}}
\newcommand{\nn}{\nonumber}
\newcommand{\bx}{{\bm x}}
\newcommand{\bk}{{\bm k}}
\newcommand{\bl}{{\bm{\ell}}}
\newcommand{\bp}{{\bm p}}
\newcommand{\bq}{{\bm q}}

\begin{document}
	\preprint{IPMU19-0135}
	\preprint{YITP-19-88}
	\title{Universal infrared scaling of gravitational wave background spectra}
	\author{Rong-Gen Cai${}^{a,b}$,~Shi Pi${}^{a,c}$ and Misao Sasaki${}^{c,a,d,e}$\\
		\it
		$^{a}$CAS Key Laboratory of Theoretical Physics, \\
		Institute of Theoretical Physics,
		Chinese Academy of Sciences, Beijing 100190, China\\
		$^{b}$School of Physical Sciences,
		University of Chinese Academy of Sciences, Beijing 100049, China\\
		$^{c}$Kavli Institute for the Physics and Mathematics of the Universe (WPI), Chiba 277-8583, Japan\\
		$^{d}$Yukawa Institute for Theoretical Physics, Kyoto University, Kyoto 606-8502, Japan\\
		$^{e}$Leung Center for Cosmology and Particle Astrophysics,National Taiwan University, Taipei 10617}   
	\date{\today}
	\begin{abstract}
		We study the general infrared behavior of the power spectrum of a stochastic gravitational wave background produced by stress tensor in the form bilinear in certain dynamical degrees of freedom.
		We find $\Omega_{\text{GW}}\propto k^3$ for a very wide class of the sources which satisfy
		a set of reasonable conditions. Namely, the $k^3$ scaling is universally valid when the source term is 
		bounded in both frequency and time, is effective  in a radiation-dominated stage, and
		for $k$ smaller than all the physical scales 
		associated with the source, like the peak frequency, peak width, and time duration, etc.
		We also discuss possible violations of these conditions and their physical implications. 
\end{abstract}
	\maketitle

\noindent	
\textit{Introduction.}~
The discovery of gravitational waves (GWs) from mergers of binary black holes (BBHs) 
and binary neutron stars (BNSs) by LIGO/VIRGO~\cite{Abbott:2016blz,Abbott:2016nmj,Abbott:2017vtc,Abbott:2017gyy,Abbott:2017oio,TheLIGOScientific:2017qsa} has brought the dawn to GW cosmology.  Once produced, GWs propagate almost freely in the universe, they  therefore carry the information about their origins as well as
the evolution of the universe. The stochastic gravitational wave backgrounds (SGWB)
may originate from many different physical sources like BH/NS binaries~\cite{Mandic:2016lcn,Clesse:2016ajp,Wang:2016ana,Garcia-Bellido:2017aan,Guo:2017njn,Raidal:2017mfl}, first order phase transitions during the evolution of the universe~\cite{Witten:1984rs,Hogan:1986qda,Apreda:2001us,Grojean:2006bp,Huang:2016odd,Cai:2017tmh,Chao:2017vrq,Wan:2018udw}, 
spectator field(s)~\cite{Bartolo:2007vp,Biagetti:2013kwa,Cai:2019jah,Biagetti:2014asa,Fujita:2014oba}, 
reheating/preheating after inflation~\cite{Khlebnikov:1997di,GarciaBellido:1998gm,Tashiro:2003qp,Easther:2006vd,GarciaBellido:2007dg,GarciaBellido:2007af,Dufaux:2007pt,Easther:2007vj,Price:2008hq,Jedamzik:2010dq,Jedamzik:2010hq,Kuroyanagi:2015esa,Liu:2017hua,Kuroyanagi:2017kfx,Cai:2018tuh,Amin:2018xfe,Liu:2018rrt}, topological defects~\cite{Figueroa:2012kw,Binetruy:2012ze,Kawasaki:2011vv,Hiramatsu:2013qaa,Gleiser:1998na,Liu:2019lul}, primordial magnetic field~\cite{Durrer:1999bk,Pogosian:2001np,Caprini:2003vc,Caprini:2009pr,Shaw:2009nf,Saga:2018ont,Caprini:2001nb,Caprini:2006jb}, and primordial scalar and tensor perturbations from inflation. 

Different GW experiments are sensitive to different  frequencies. GWs with cosmological wavelengths,
i.e., frequencies of order $10^{-16}$ Hz,
can be indirectly detected by the B-mode polarization of the cosmic  microwave background (CMB)~\cite{Ade:2015tva,Li:2017drr,Matsumura:2013aja}. 
For GWs with frequency of order $10^{-9}$Hz, pulsar timing array (PTA) is the most effective detector~\cite{Sazhin:1977tq,Detweiler:1979wn,Desvignes:2016yex,Hobbs:2013aka,McLaughlin:2013ira,Verbiest:2016vem}. LIGO and LIGO-like interferometers can detect GWs of $10\sim 10^3~\text{Hz}$~\cite{Aasi:2013wya,Punturo:2010zz,Sathyaprakash:2012jk}, while space-based interferometers like LISA~\cite{AmaroSeoane:2012km,AmaroSeoane:2012je,Audley:2017drz}, Taiji~~\cite{Guo:2018npi}, Tianqin~\cite{Luo:2015ght}, Decigo~\cite{Kawamura:2011zz},
 and beyond~\cite{Crowder:2005nr,Corbin:2005ny,Baker:2019pnp} 
 are sensitive to smaller frequencies $10^{-4}\sim1~\text{Hz}$.  

Cosmologists describe the  SGWBs by their energy density per logarithmic frequency normalized by the current critical energy density of the universe, $\Omega_\text{GW}$. Most of the GW signals generated by given physical processes exhibit a power-law structure on the red
 side of a characteristic frequency $f_*$, e.g., $\Omega_\text{GW}\propto f^\beta$ for $f<f_*$. 
Searching and identification of a SGWB spectrum also rely on the ansatz of such a power-law scaling~\cite{Kuroyanagi:2018csn,Caprini:2019pxz}. 
For instance, an incoherent superposition of GWs emitted by BBHs has a characteristic scaling $\beta=2/3$~\cite{Phinney:2001di,Regimbau:2011rp,Zhu:2011bd,Rosado:2011kv,Marassi:2011si,Zhu:2012xw}. Secondary GWs induced by scalar curvature perturbations with a broad peak scales as $f^3$, while a $\delta$-function peak gives $f^2$. For SGWB from first order phase transitions, Ref.\cite{Caprini:2009fx} has noticed the universal $f^3$-scaling for small $f$, and justified it by a causality argument. But some steeper powers are observed later, especially for long-lived highly oscillating sources~\cite{Hindmarsh:2016lnk,Hindmarsh:2017gnf,Inomata:2019zqy,Inomata:2019ivs,Hindmarsh:2019phv}.
Therefore it is interesting to study the infrared scaling of SGWB spectra in a general way 
that can be applied to a wide class of sources, and find out under which conditions 
the $k^3$-law may be obtained. This is the main task of this work.
\\

\noindent
\textit{Gravitational Waves from Bilinear Source.}~
The metric we assume is
\be
ds^2=a^2(\eta)(-d\eta^2+(\delta_{ij}+h_{ij})dx^idx^j),
\ee
where $h_{ij}(\eta,\bx)$ is a transverse traceless (i.e., tensor) perturbation.
We expand it as
\be
h_{ij}(\eta,\bx)=\int\frac{d^3k}{(2\pi)^{3/2}}\sum_{\lambda=+,\times}e_{ij}^\lambda(\hat k)h_{\bk,\lambda}(\eta)e^{i\bk\cdot\bx},
\ee
where $e_{ij}^{+,\times}(\hat k)$ are two orthonormal polarization tensors of GWs perpendicular to the $\hat k$ direction with $e^\lambda_{ij}e_\mu^{ij}=\delta^\lambda_\mu$ and $\sum_\lambda e_{ij}^\lambda e^{lm}_\lambda=\Lambda_{ij}^{lm}$. $\Lambda_{ij}^{lm}$ is the transverse traceless projector defined as $\Lambda_{ij}^{lm}\equiv\frac12\left(\pi^l_i\pi_j^m+\pi^l_j\pi_i^m-\pi_{ij}\pi^{lm}\right)$, 
where $\pi_{ij}\equiv\delta_{ij}-\hat k_i\hat k_j$ is the transverse projector perpendicular to the $\hat k$ direction. 
It is customary to describe the SGWB by the energy density of GWs per logarithmic frequency 
normalized by the critical density,
 $\Omega_\text{GW}\equiv(\rho_\text{crit})^{-1}d\rho_\text{GW}/d\ln k$, 
 where $\rho_\text{crit}=3H^2/(8\pi G)$ and 
\begin{align}\nn
\rho_\text{GW}
&=\frac{1}{64\pi Ga^2}\int\frac{dk}{k}\frac{k^3}{2\pi^2}\\\nn
&~~~\sum_{\lambda=+,\times}\langle h'_{\bk,\lambda}(\eta)h_{\bp,\lambda}^*{}'(\eta)
+k^2h_{\bk,\lambda}(\eta)h_{\bp,\lambda}^*(\eta)\rangle'_{\bp=\bk}.
\end{align}
where a prime on $h_\bk$ denotes the derivative with respect to the conformal time $\eta$,
and a prime on an angular bracket means an overall $\delta^{(3)}(\bk-\bp)$ factor is removed. 
The angular brackets mean an ensemble average. 
We start from the equation of motion for the Fourier mode $h_\bk$,
\be\label{eom:hij}
h_{\bk,\lambda}''+2\mathcal{H}h_{\bk,\lambda}'+k^2h_{\bk,\lambda}=16\pi Ga^2e^{ij}_\lambda\Lambda_{ij}^{ab}(\hat{k})T_{ab,\bk},
\ee
where $\mathcal{H}\equiv a'/a=aH$ is the conformal Hubble parameter, and $T_{ab}$ is the spatial components of the energy-momentum tensor whose transverse traceless part sources the tensor perturbation. We will discuss the origins of $T_{ab}$ later. 
The solution to \eqref{eom:hij} can be written as~\cite{Ananda:2006af}
\be\label{sol:h}
h_{\bk,\lambda}(\eta)=\frac2{a(\eta)\mpl^2}\int^\infty_0\!\!\!\!d\tilde\eta~a^3(\tilde\eta)
G_k(\eta;\tilde\eta)e_\lambda^{ij}\Lambda^{ab}_{ij}(\hat k)T_{ab,\bk}(\tilde\eta),
\ee
where $G_k(\eta;\tilde\eta)$ is the retarded Green function of the equation, $v_\bk''+(k^2-a''/a)v_\bk=\delta(\eta-\tilde\eta)$ with $v_\bk=ah_\bk$, 
which in the radiation-dominated universe takes the form,
\be
G_k(\eta;\tilde\eta)=\frac{\sin k(\eta-\tilde\eta)}{k}\Theta(\eta-\tilde\eta),
\ee
where $\Theta$ is the Heaviside step function. 
 The GW energy density parameter is
\begin{align}\nn
&\Omega_\text{GW}(k)=\frac{k^3}{12\pi^2a^4H^2\mpl^4}\int_0^\eta d\eta_1\int_0^\eta d\eta_2~a^3(\eta_1)a^3(\eta_2)\\\label{def:OmegaGW}
&~\times\cos k(\eta_1-\eta_2)
\Lambda_{ab}^{cd}(\hat{k})\langle T^{ab}_\bk(\eta_1)T^*_{cd,\bp}(\eta_2)\rangle'_{\bp=\bk}\,.
\end{align}
where a prime on the correlator means omitting an overall $\delta$ function $\delta^{(3)}(\bk-\bp)$
as before. 

The energy-momentum tensor that sources GWs is definitely model dependent. 
However, for a broad class of models, it is in the \textit{bilinear} form,
\be\label{def:Tij}
T_{ab}(\eta,\bx)=\partial_a\phi(\eta,\bx)\partial_b\phi(\eta,\bx)
+v_{a}(\eta,\bx)v_{b}(\eta,\bx),
\ee
up to a total derivative, where $\phi$ is a scalar field, and $v_a$ is a vector field 
which can be decomposed into the divergence and transverse parts 
as $v_a=\partial_av+w_a$. 
This is the most general form of the stress tensor in the form bilinear in 
scalar or vector degrees of freedom. After transforming it to momentum space, 
\eqref{def:Tij} becomes a sum of convolutions, with its two-point function given by
\begin{align}\nn
&\langle T^{ab}_\bk(\eta_1)T^{cd*}_{\bp}(\eta_2)\rangle\\\nn
&=\int\frac{d^3ld^3q}{(2\pi)^3}\Big[\langle v_{\bl}^{a}(\eta_1)v_{\bk-\bl}^{b}(\eta_1)v^{c*}_{\bq}(\eta_2)v^{d*}_{\bp-\bq}(\eta_2)\rangle\\\label{<TT>}
&+\bl^a(\bk-\bl)^b\bq^c(\bp-\bq)^d
\langle\phi_{\bl}(\eta_1)\phi_{\bk-\bl}(\eta_1)\phi^*_{\bq}(\eta_2)\phi^*_{\bp-\bq}(\eta_2)\rangle\Big].
\end{align}
The four-point function of the vector $v^a$ may be expressed as 
\begin{align}\nn
&
\big\langle v^a_\bl(\eta_1)v^b_{\bk-\bl}(\eta_1)v^{c*}_\bq(\eta_2)v^{d*}_{\bp-\bq}(\eta_2)\big\rangle\\\nn
&=\big\langle v^a_\bl(\eta_1)v^{c*}_\bq(\eta_2)\big\rangle\big\langle v^b_{\bk-\bl}(\eta_1)v^{d*}_{\bq-\bp}(\eta_2)\big\rangle\\\nn
&+\big\langle v^a_\bl(\eta_1)v^{d*}_{\bp-\bq}(\eta_2)\big\rangle\big\langle v^b_{\bk-\bl}(\eta_1)v^{c*}_\bq(\eta_2)\big\rangle\\\label{wick}
&+\big\langle v^a_\bl(\eta_1)v^b_{\bk-\bl}(\eta_1)v^{c*}_\bq(\eta_2)v^{d*}_{\bp-\bq}(\eta_2)\big\rangle_c,
\end{align}
where the contribution of the connected four-point function to the GW energy density will vanish 
by symmetry. A similar expression holds for the scalar $\phi$. 
The two-point function $\langle v^a_\bl v^{c*}_\bq\rangle$ can be decomposed 
into the parts longitudinal and perpendicular to the $\bl$-direction, 
while the two-point function of the scalar $\langle\phi_\bl\phi_\bq^*\rangle$ is as usual,
\begin{align}\nn
&\big\langle v^{a}_\bl(\eta_1)v^{c*}_\bq(\eta_2)\big\rangle=\delta^{(3)}(\bl-\bq)\frac{2\pi^2}{\ell^3}\\\label{<vv>}
&\times\ell^2
\Big[\mathcal{P}_w(\eta_1,\eta_2,l)\pi^{ac}(\bl)+\mathcal{P}_v(\eta_1,\eta_2,l)\hat\ell^a\hat\ell^c\Big],
\\\label{<phiphi>}
&\langle\phi_{\bl}(\eta_1)\phi^*_{\bq}(\eta_2)\big\rangle=\delta^{(3)}(\bl-\bq)\frac{2\pi^2}{\ell^3}
\mathcal{P}_\phi(\eta_1,\eta_2,\ell),
\end{align}
where $\pi^{ac}(\bl)=\delta^{ab}-\hat \ell^a\hat \ell^c$. 
Note that since \eqref{<vv>} and \eqref{<phiphi>} are unequal-time correlators, $\mathcal{P}_\phi$, $\mathcal{P}_v$, and $\mathcal{P}_w$ may not be positive definite unless $\eta_1=\eta_2$.
As $v_a$ and $\phi$ are independent variables, we  have
\begin{align}\label{<vv><vv>}
&\quad\big\langle v^{a}_\bl(\eta_1)v^{c*}_{\bq}(\eta_2)\big\rangle\big\langle v^{b}_{\bk-\bl}(\eta_1)v^{d*}_{\bp-\bq}(\eta_2)\big\rangle\\\nn
&=\delta^{(3)}(\bl-\bq)\delta^{(3)}(\bk-\bp)\frac{4\pi^4}{\ell|\bk-\bl|}\\\nn
&\Big(\mathcal{P}_w(\ell)\mathcal{P}_w(|\bk-\bl|)\pi^{a}_{c}(\hat\ell)\pi^{b}_{d}(\hat n)+\mathcal{P}_v(\ell)\mathcal{P}_w(|\bk-\bl|)\hat{\ell}^{a}\hat{\ell}_{c}\pi^{b}_{d}(\hat n)\\\nn
&+\mathcal{P}_w(\ell)\mathcal{P}_v(|\bk-\bl|)\pi^{a}_{c}(\hat\ell)\hat{n}^{b}\hat{n}_{d}
+\mathcal{P}_v(\ell)\mathcal{P}_v(|\bk-\bl|)\hat{\ell}^{a}\hat{n}^{b}\hat{\ell}_{c}\hat{n}_{d}\Big),\\
\label{<ss><ss>}
&\langle\phi_\bl(\eta_1)\phi^*_{\bk-\bl}(\eta_2)\rangle
\langle\phi_{\bk-\bl}(\eta_1)\phi^*_{\bp-\bq}(\eta_2)\rangle\\\nn
&=\delta^{(3)}(\bl-\bq)\delta^{(3)}(\bk-\bp)\frac{4\pi^4}{l^3|\bk-\bl|^3}
\mathcal{P}_\phi(\ell)\mathcal{P}_\phi(|\bk-\bl|).
\end{align}
where $\hat{\ell}\equiv\bl/\ell$ and $\hat{n}\equiv(\bk-\bl)/|\bk-\bl|$. 
In the infrared limit, $k$ is smaller than any scale in the source, thus $\hat n\rightarrow-\hat \ell$. 
This is equivalent to picking up the leading order in the multipole expansion around $\bk$ 
in \eqref{<vv><vv>} and \eqref{<ss><ss>}. 
As there is no $k$ dependence in the leading order, we can expand the transverse 
projectors to combinations of Kronecker $\delta$ and unit vector $\hat\ell$, 
\begin{align}\nn
&\quad\big\langle v^{a}_\bl(\eta)v^{c*}_{\bq}(\tau)\big\rangle\big\langle v^{b}_{\bk-\bl}(\eta)v^{d*}_{\bp-\bq}(\tau)\big\rangle\\\nn
&\to\delta^{(3)}(\bl-\bq)\delta^{(3)}(\bk-\bp)\frac{4\pi^4}{\ell^2}
\Big[\left(\mathcal{P}_w-\mathcal{P}_v^2\right)^2\hat{\ell}^{a}\hat{\ell}^{b}\hat{\ell}_{c}\hat{\ell}_{d}
\\\label{<vv><vv>'}
&+\left(\mathcal{P}_v\mathcal{P}_w-\mathcal{P}_w^2\right)\hat{\ell}^{a}\hat{\ell}_{c}\delta^{b}_{d}
+\left(\mathcal{P}_w\mathcal{P}_v-\mathcal{P}_w^2\right)\delta^{a}_{c}\hat{\ell}^{b}\hat{\ell}_{d}
+\mathcal{P}_w^2\delta^{a}_{c}\delta^{b}_{d}
\Big],\\\nn
&\bl^a(\bk-\bl)^b\bq^c(\bp-\bq)^d\langle\phi_\bl(\eta_1)\phi^*_{\bk-\bl}(\eta_2)\rangle
\langle\phi_{\bk-\bl}(\eta_1)\phi^*_{\bp-\bq}(\eta_2)\rangle\\\label{<ss><ss>'}
&\longrightarrow\delta^{(3)}(\bl-\bq)\delta^{(3)}(\bk-\bp)\frac{4\pi^4}{\ell^2}
\mathcal{P}_\phi^2\hat{\ell}^a\hat{\ell}^b\hat{\ell}^c\hat{\ell}^d.
\end{align}
Here the arguments of the functions $\mathcal{P}_{\phi,v,w}(\eta_1,\eta_2,l)$ are not explicitly written. 
Then we substitute \eqref{<vv><vv>'} and \eqref{<ss><ss>'} into \eqref{<TT>} and \eqref{wick}. 
The overall $\delta$ function, $\delta^{(3)}(\bk-\bl)$, will be kept until the end, while $\delta^{(3)}(\bl-\mathbf{q})$ can be integrated by $\int d^3q$ in \eqref{<TT>}, making $\bq\rightarrow\bl$. The integral over $d^3\ell$ can be written as $\ell^2d\ell d\Omega_\ell$, 
which can greatly simplify the multipole moments as
\begin{align}
\int d\Omega_\ell~\hat \ell_i\hat \ell_j&=\frac{4\pi}{3}\delta_{ij},\\
\int d\Omega_\ell~\hat \ell_i\hat \ell_j\hat \ell_\ell\hat\ell_m
&=\frac{4\pi}{15}\left(\delta_{ij}\delta_{\ell m}+\delta_{i\ell}\delta_{jm}+\delta_{im}\delta_{j\ell}\right).
\end{align}
All these results will be contracted with $\Lambda_{ab}^{cd}$ in \eqref{def:OmegaGW}. 
By using the traceless property of $\Lambda$ as well as $\Lambda^{ab}_{ab}=2$, we have
\begin{align}\nn
&\Omega_\text{GW}(\eta,k)=\frac{k^3}{45a^4H^2\mpl^4}\int_0^\eta d\eta_1\int_0^\eta d\eta_2\,a^3(\eta_1)a^3(\eta_2)\\\label{OmegaGW1}
&\times\cos k(\eta_1-\eta_2)\int d\ell
\left[\left(2\mathcal{P}_v+3\mathcal{P}_w\right)^2+5\mathcal{P}_w^2+4\mathcal{P}_\phi^2\right].
\end{align}
The solution \eqref{sol:h} is valid only in  the radiation-dominated universe, so \eqref{OmegaGW1} 
holds only until the matter-radiation equality, $\eta<\eta_{\rm eq}$. 
For a wide class of models, the source exists only for a finite duration of time $\Delta\eta_s$. 
When going to the small $k$ limit, 
 we have $k\ll\Delta\eta_s^{-1}$, and the cosine function can be approximately taken to be 1. 
 Then taking account of the redshift factor from the equal time to present,
we obtain
\begin{align}\nn
&\Omega_\text{GW}(\eta_0,k)\\\nn
&=
\frac{k^3}{45a_0a_\text{eq}^3H_\text{eq}^2\mpl^4}\int_0^{\eta_\text{eq}}d\eta_1\int_0^{\eta_\text{eq}}d\eta_2\,a^3(\eta_1)a^3(\eta_2)
\\\label{result:OmegaGW}
&\quad\times\int d\ell
\left[\left(2\mathcal{P}_v+3\mathcal{P}_w\right)^2+5\mathcal{P}_w^2+4\mathcal{P}_\phi^2\right].
\end{align}
Note that the quadratic forms in \eqref{result:OmegaGW} are always positive.
Thus assuming the integral is finite, we have
\be\label{powerint}
0<\int d\ell
\left[\left(2\mathcal{P}_v+3\mathcal{P}_w\right)^2+5\mathcal{P}_w^2+4\mathcal{P}_\phi^2\right]<\infty.
\ee
Since this integral is $k$-independent, we find the universal scaling 
$\Omega_{\rm GW}(\eta_0,k)\propto k^3$ in the limit $k\to0$.

To summarize, the above can be stated as the following theorem: 
The GW spectrum we observe today scales as
$\Omega_\text{GW}(\eta_0,k)\propto k^3$ in the infrared regime
if the source satisfies the conditions:
	\begin{enumerate}[label=(\Roman*)]
	\item The integral \eqref{powerint} is finite.\label{c1}
	\item $k$ is smaller than all the scales associated with the source term, for instance $k_*$ (characteristic frequency), $\Delta k$ (peak width), and $\Delta t_s^{-1}\equiv(a_*\Delta\eta_s)^{-1}$ (duration of the source).\label{c2}
	\item Modes of interest reenter the Hubble horizon during the radiation-dominated era. \label{c3}
	\end{enumerate}
We note that $k$ is connected to the GW frequency today by $f=ck/(2\pi a_0)$.

Let us list below a few specific examples of the sources in the bilinear form~\eqref{def:Tij}, 
and see if they satisfy the above conditions and give the $k^3$ scaling. 
The violation of these conditions will be discussed later.
\\

\noindent
(a) {\it Secondary GWs induced by scalar perturbations}~
\cite{Ananda:2006af,Baumann:2007zm,Alabidi:2012ex,Alabidi:2013wtp,Inomata:2016rbd,Orlofsky:2016vbd,Kohri:2018awv,Assadullahi:2009nf,Assadullahi:2009jc,Gong:2017qlj,Espinosa:2018eve,Cai:2018dig,Bartolo:2018rku,Byrnes:2018txb,Inomata:2018epa,Dalianis:2018ymb,Cai:2019amo,Wang:2019kaf,DeLuca:2019qsy,Tada:2019amh,Unal:2018yaa,Xu:2019bdp,Yuan:2019udt,Saito:2008jc,Saito:2009jt,Bugaev:2009zh,Bugaev:2010bb,Nakama:2016gzw,Cai:2019elf,Lu:2019sti,Kalaja:2019uju,Chen:2019zza,Atal:2019erb,Bartolo:2018evs,Gong:2019mui}. 
Tensor perturbations are decoupled from the scalar perturbation at linear order, 
but they are coupled at second order. 
Such induced secondary GWs are negligible on CMB scales as the curvature perturbation $\mathcal R$ is highly constrained to be of order $10^{-5}$~\cite{Akrami:2018odb}. 
However, the power spectrum of the curvature perturbation may have peak(s) on small scales 
and it is suggested that a substantial amount of primordial black holes (PBHs) may
form when the rare, very high peaks reenter the horizon~\cite{Zeldovich:1963,Hawking:1971ei,Carr:1974nx,Meszaros:1974tb,Carr:1975qj,Sasaki:2018dmp}, 
which, according to their constraints of different masses~\cite{Frampton:2009nx,Carr:2009jm,Carr:2016drx,Green:2004wb,Poulter:2019ooo,Young:2014ana,Byrnes:2018clq,Tisserand:2006zx,Graham:2015apa,Koushiappas:2017chw,Authors:2019qbw,Niikura:2017zjd,Katz:2018zrn,Montero-Camacho:2019jte,Carr:2016hva}, can be a candidate of dark matter~\cite{GarciaBellido:1996qt,Kawasaki:1997ju,Yokoyama:1998pt,Kohri:2012yw,Clesse:2015wea,Inomata:2017okj,Garcia-Bellido:2017mdw,Kannike:2017bxn,Inomata:2017vxo,Ando:2017veq,Ando:2018nge,Espinosa:2017sgp,Cheng:2016qzb,Cheng:2018yyr,Frampton:2010sw,Kawasaki:2012wr,Pi:2017gih}, 
the merging black holes detected by LIGO/VIRGO~\cite{Bird:2016dcv,Sasaki:2016jop,Chen:2016pud,Wang:2016ana,Blinnikov:2016bxu,Ali-Haimoud:2016mbv,Garcia-Bellido:2017imq,Guo:2017njn,Zumalacarregui:2017qqd,Clesse:2016vqa,Garcia-Bellido:2017aan,Mandic:2016lcn,Raidal:2017mfl,Clesse:2016ajp}, or seeds for structure formation~\cite{Kawasaki:2012kn,Bean:2002kx,Nakama:2017xvq,Carr:2018rid,Nakama:2019htb}.
A simple estimate gives that the density parameter of GWs generated from the second-order 
scalar perturbations are roughly of order $\mathcal R^2$, where $\cal R$ is the amplitude of
the conserved comoving curvature perturbation on superhorizon scales. 
The GW spectrum peaks at the characteristic frequency proportional to 
$M_\text{PBH}^{-1/2}$, and may have unique features around the peak. 
Neglecting the isotropic components, the source stress tensor takes the form, 
\be\label{Tij_IGW}
\frac{T_{ij}}{\mpl^{2}}
=-2\Phi\frac{\partial_i\partial_j}{a^2}\Phi+\frac{\partial_i}{a}\left(\Phi+\frac{\Phi'}{\mathcal H}\right)\frac{\partial_j}{a}\left(\Phi+\frac{\Phi'}{\mathcal{H}}\right)
\ee
where $\Phi$ is the curvature perturbation on the Newton slices. 
During the radiation-domination,
$\Phi$ is given by
\begin{align}
\Phi_\bl=2{\cal R}(\bl)\frac{j_1(c_s\ell\eta)}{c_s\ell\eta}\,,
\label{Phirad}
\end{align}
where ${\cal R}(\bl)$ is the conserved comoving curvature perturbation on superhorizon scales
and $c_s=1/\sqrt{3}$.
When the scale enters the horizon, $\ell\eta\gg1$,
 the source term $\Phi_\bl(\eta)$ decays rapidly as $(\ell\eta)^{-2}$.
Thus the $(\partial_i\Phi'\partial_j\Phi')$ term dominates on subhorizon scales. 
Then when the power spectrum of $\Phi$ has a peak at a certain scale $k_*$ with 
 a width $\sigma$, which is usually the case for PBH formation scenarios,  
 the finiteness of the integral \eqref{powerint} is assured.
 We thus reach the conclusion that $\Omega_\text{GW}\propto k^3$ for $k\ll\text{min}(k_*,\sigma)$. 
It is impossible for this condition to be satisfied
 when the width is infinitesimally small, i.e. $\sigma\rightarrow0$ for the
  $\delta$-function peak, which we will discuss later.
\\

\noindent
(b) {\it GWs from first-order phase transitions}~\cite{Witten:1984rs,Hogan:1986qda}.
A first-order phase transition may take place in various particle physics models~\cite{Apreda:2001us,Grojean:2006bp,Cai:2017tmh,Chao:2017vrq,Huang:2016odd,Wan:2018udw}. When it happens, bubbles nucleate at a rate $\Gamma(t)$. 
The bubbles expand and collide after some characteristic time 
duration $\Delta t\approx1/\beta\approx\Gamma/\dot\Gamma$.  
So the characteristic radius at collision is $R_*=v_w/\beta$, where $v_w$ is the velocity
 of the bubble wall in the rest frame of the bubble center. 
 Besides the bubble collisions ~\cite{Turner:1990rc,Kosowsky:1991ua,Kosowsky:1992rz,Kosowsky:1992vn,Turner:1992tz,Kamionkowski:1993fg,Caprini:2007xq,Jinno:2016vai,Cutting:2018tjt}, 
 there are also compressible modes (sound waves)
 ~\cite{Hindmarsh:2013xza,Hindmarsh:2017gnf,Hindmarsh:2016lnk,Hindmarsh:2015qta} 
 and the following magnetohydrodynamic (MHD) turbulence
 ~\cite{Kamionkowski:1993fg,Kahniashvili:2008pe,Caprini:2009yp,Kosowsky:2001xp,Dolgov:2002ra,Kahniashvili:2005qi,Gogoberidze:2007an}, 
 all of which can generate GWs by the relevant part of the source term 
\be
T_{ij}=\partial_i\phi\partial_j\phi+\frac{\left(\rho+p\right)V_iV_j}{1-V^2},
\ee
where $\phi$ is the scalar field that triggers the phase transition, 
and $V_i$ is the velocity field of the bubble wall or the fluid in the rest frame of bubble center. MHD turbulence may also enhance the magnetic field, which also sources the GWs, as
\be
T_{ij}=\frac1{4\pi}\left[B_iB_j-\frac12\delta_{ij}B^2\right].
\ee
Among these three processes, bubbles collide in a short period of time $\Delta\eta_\text{coll}=a_*/\beta$, while the sound waves and/or turbulence may exist for a longer period of time $\Delta\eta_\text{sw}$, $\Delta\eta_\text{turb}$. The GW spectrum scales as $\Omega_\text{GW}\propto k^3$ for $k\ll \text{min}(a_*/R_*, \Delta\eta_\text{sw}^{-1}, \Delta\eta_\text{turb}^{-1})$. There will be some subtlety when $\Delta\eta_\text{sw}$ or $\Delta\eta_\text{turb}$ is larger than $a_*/R_*$, which we will discuss later. 
The magnetic field itself can induce GWs, which is highly constrained by the CMB B-mode polarization~\cite{Durrer:1999bk,Pogosian:2001np,Caprini:2003vc,Caprini:2009pr,Shaw:2009nf,Saga:2018ont,Caprini:2001nb,Caprini:2006jb}.
\\

\noindent
(c)
{\it GWs induced by an inflaton or a spectator scalar field in preheating after inflation}
~\cite{Tashiro:2003qp,Easther:2006vd,GarciaBellido:2007dg,Dufaux:2007pt,Kuroyanagi:2015esa,Liu:2017hua,Kuroyanagi:2017kfx,Cai:2018tuh,Jedamzik:2010dq,Amin:2018xfe,Jedamzik:2010hq,Khlebnikov:1997di,GarciaBellido:1998gm,Easther:2007vj,GarciaBellido:2007af,Price:2008hq}. The source term is $T_{ij}=\partial_i\phi\partial_j\phi$, where $\phi$ is the inflaton or some other intermediate scalar field.
The parametric resonance may largely enhance the amplitude of $\phi$, which in turn may
generate substantial GWs. Detailed study should be done by lattice simulations, 
but the infrared scaling is $k^3$ as expected, as was first pointed out in \cite{Easther:2006vd}.
 An related case is GWs from domain walls~\cite{Kawasaki:2011vv,Hiramatsu:2013qaa,Gleiser:1998na}. 
\\

\noindent
\textit{Violation of the conditions.}~
In some processes, one or more of the three conditions we listed will be violated, and some non-cubed infrared scaling will appear. Condition \ref{c1} breaks down if \eqref{powerint} diverges or vanishes. Let us first consider a power-law divergence $k^{-\gamma}$ ($\gamma>0)$ of \eqref{powerint}
 in the limit $k\to0$.  This would mean $\Omega_\text{GW}(\eta_0)\propto k^{3-\gamma}$. 
  In passing, it is worth mentioning that in this case of the power-law divergence,
  the power spectrum index of $\Omega_\text{GW}(\eta_0)$ will be always smaller than 3. 

For instance, SGWB induced by scale-invariant primordial scalar perturbations is also scale-invariant.
In this case, ${\cal P}_\Phi^2$ in \eqref{result:OmegaGW} is proportional to $l^{-4}$, which would give rise to a cubic divergence of the integral. In other words, the integral would behave
 as $k^{-3}$ in the $k\rightarrow0$ limit, thus cancel the $k^3$ coefficient in \eqref{def:OmegaGW}, 
 resulting in a scale-invariant $\Omega_{\rm GW}(k,\eta_0)$.
Another example is the SGWB induced by scalar perturbations with a $\delta$-function
spectrum, $\mathcal{P}_\Phi=\mathcal{A}\delta(\ln(\ell/k_*))$,
where \eqref{result:OmegaGW} displays a linear divergence thus is proportional to $1/k$ when $k\rightarrow0$. 
The resultant spectrum is expected to behave as $k^{3-1}=k^2$, 
as is shown in Refs.\cite{Ananda:2006af,Saito:2008jc,Alabidi:2012ex,Alabidi:2013wtp,Orlofsky:2016vbd,Kohri:2018awv,Byrnes:2018txb,Dalianis:2018ymb,Cai:2019amo,Inomata:2016rbd,Wang:2019kaf}.
We note that, in addition to the divergence of integral \eqref{result:OmegaGW}, as the $\delta$ function has a zero width, Condition \ref{c2} is also violated. 
We would like to comment that $\delta$-function peak is unphysical, but only convenient for reaching an analytical expression of $\Omega_\text{GW}$.

In some physical processes the source term as well as $\mathcal{P}(\eta_1,\eta_2,l)$ highly oscillates 
and the oscillations resonate with $\cos k(\eta_1-\eta_2)$.
In this case, we cannot directly take the $k\rightarrow0$ limit in \eqref{OmegaGW1}, 
as the main contributions to integral  \eqref{OmegaGW1} come from the resonances. 
The behavior near the resonance point will render the resulting $k$-dependent, making it completely model-dependent.

One example is the SGWB from the incoherent superpositions
 of the compact binaries. 
Every binary is an oscillating system with time-dependent frequency 
$\omega_\text{s}$, which gives the resonance frequency of integral \eqref{OmegaGW1} $k=2\omega_\text{s}$.
The simple infrared limit $k\rightarrow0$ in \eqref{OmegaGW1} becomes meaningless.
The time integral in \eqref{OmegaGW1} can be easily evaluated by the stationary 
 phase approximation, which gives the correct scaling 
  $\Omega_{\text{GW}}\propto k^{2/3}$~\cite{Phinney:2001di,Regimbau:2011rp,Zhu:2011bd,Rosado:2011kv,Marassi:2011si,Zhu:2012xw}. The $k^3$ scaling is, however, still valid for scales larger than the maximum orbital radius of
 the binaries, at their formation epoch. 
 This corresponds to a current frequency 
$f\ll a_\text{eq}(\rho_\text{eq}/M_\text{BH})^{1/3}\approx(M_\odot/M_{BH})^{1/3}
\times10^{-10}~\text{Hz}$~\cite{Nakamura:1997sm} for PBH binaries, 
which is far smaller than the detectable band of any experiment. 

Another example is the sound waves and the MHD 
turbulence of the first-order phase transition, provided that their duration $\Delta\eta$ is
longer than the characteristic time of bubble collision. 
In such a case, the source term can be decomposed into harmonics with freqency $c_s\ell$ inside
 the horizon, thus the main contribution to integral \eqref{OmegaGW1} comes from $k\sim c_s\ell$, 
 which gives a steep GW spectrum scales at most as $\Omega_\text{GW}\propto k^9$
for $\Delta\eta^{-1}\ll k\ll a_*/R_*$~\cite{Hindmarsh:2016lnk,Hindmarsh:2019phv}. 
Nevertheless, for $k\ll \Delta\eta^{-1}$, the $k^3$ law still holds. 
Note that when $\Delta\eta$ is larger than 
the Hubble horizon scale at the collision time, $(a_*H_*)^{-1}$, 
the duration time $\Delta\eta$ is approximately equal to 
the Hubble horizon scale when the source ceases to exist,
i.e., $\Delta\eta\simeq(a_\text{e}H_\text{e})^{-1}$.
\\

\textit{Conclusion.}~
In this paper we presented a general expression and studied the spectrum of 
stochastic GWs generated by the energy-momentum tensor bilinear in the source, 
and showed that the infrared scaling of the GW spectrum $\Omega_\text{GW}$ behaves
as $k^3$ for a wide class of models, i.e., for those sources whose power spectra are
essentially localized in the Fourier space as well as in time domain, and the modes of interest
enter the Hubble horizon in the radiation-dominated stage.
In particular, the $k^3$ scaling is found to be always valid on superhorizon scales 
where the source disappears ($k\ll(a_\text{e}H_\text{e})^{-1}$). 
This can be explained by causality~\cite{Suyama:2004mz,Linde:1996gt,Liddle:1999hq}: 
Suppose the source of the GWs peaks at a characteristic scale $1/k_*$.
 For an infrared scale $1/k>1/k_*$, at the horizon crossing of $1/k_*$, there are $N=(k_*/k)^3$ 
causally disconnected patches 
where the amplitude of the tensor perturbation is $h_{*}$. As a random variable, $h_{*(i)}$ 
obeys the Poisson distribution, which means on scale of $1/k$, $h_k=\sum_ih_{*(i)}/N$. 
The two-point function of the tensor perturbation at the formation is then $\langle h_kh_k\rangle=N^{-2}\sum_{ij}\langle h_{*(i)}h_{*(j)}\rangle=N^{-1}|h_*|^2=(k/k_*)^3|h_*|^2$. $h_{k}$ stays a constant on superhorizon scales, but decays as $1/a$ after horizon crossing. It evolves to $h_k(\eta_0)$ at present by a redshift factor $a_{k}/a_0\sim1/k$, which is canceled by the $k$ factor in the definition of the energy density of GWs:
$\rho_\text{GW}\sim\langle\dot h_{k}^2\rangle\sim(k/a)^2\langle h_{k}^2\rangle$. 
Thus we obtain the scaling $\Omega_\text{GW}\propto\rho_\text{GW}\propto k^3|h_{*}|^2$.

Beyond causality argument, our discussion actually showed that during the radiation-dominated era, 
$\Omega_{\text{GW}}\propto k^3$ can still hold well inside the horizon, 
as long as $k$ is smaller than all the scales of the source, especially 
the inverse time duration for which the source exists. 
The condition that integral \eqref{powerint} is positive and finite can be guaranteed 
easily if the source is spiky and transient. But there will be some subtleties if the source 
is highly oscillating for a long period of time. We also briefly discussed such physical cases 
when the integral \eqref{powerint} contains a highly oscillating part, and found that 
the resonance between Green function and the oscillating source term becomes crucial. 
$k^3$ law will be valid for those modes which are superhorizon at the moment 
when the source term disappears or stops oscillating, otherwise it will be model-dependent. We believe our result can clarify 
some confusing understanding on the infrared power of SGWB, which is helpful in the 
future search for the SGWB signals.

Finally, we comment on Condition \ref{c3}.
 It is easy to extend our discussion to constant equation of state $w=P/\rho$ for the background evolution when $h_k$ reenters Hubble horizon. 
 From the causality argument we see that the $k^3$ factor from space dimension
 and $k^2$ factor from 
the definition of $\rho_\text{GW}$ are universal, yet the redshift factor $a_k/a_0$ depends on 
the background evolution. 
Assuming both Conditions \ref{c1} and \ref{c2} hold,
with $\eta_k\sim1/k$ and $a(\eta)\sim\eta^{2/(1+3w)}$, 
we immediately obtain
\be\label{OmegaGWgeneral}
\Omega_\text{GW}(\eta_0,k)\propto k^{3+2\frac{3w-1}{3w+1}}.
\ee
If the source reenters the horizon in the 
matter dominated epoch ($w=0$), \eqref{OmegaGWgeneral} predicts $\Omega_\text{GW}\propto k$, which has been shown in Refs.\cite{Assadullahi:2009nf,Kohri:2018awv,Inomata:2019ivs,Inomata:2019zqy,Gong:2019mui}. 
An interesting case is that the power of $k$ becomes negative when $-1/3<w<-1/15$, which we leave for future studies. 
Another interesting case is that even in the radiation-dominated universe, there will be deviation from $w=1/3$ due to the sudden change of the relativistic degrees of freedom, say, in the QCD phase transition~\cite{Byrnes:2018clq}. Such transient deviation will cause a sudden flattening in $\Omega_\text{GW}\propto k^3$, but only in a very narrow frequency band. Therefore our main result will not be altered.
 

\textit{Acknowledgements}~
We thank Christian Byrnes and Teruaki Suyama for useful discussions and comments. S. P. thanks Astronomy Department, University of Science and Technology of China, for the hospitality during his visit. R. G. C. was supported by the National Natural Science Foundation of China Grants, No. 11647601, No. 11690022, No. 11821505, No. 11851302 and No.11991052, and by the Strategic Priority Research Program of CAS Grant No. XDB23030100, and by the Key Research Program of Frontier Sciences of CAS. S.P. was supported in part by JSPS Grant-in-Aid for Early-Career Scientists No. 20K14461. M.S. was supported in part by the JSPS KAKENHI Nos. 19H01895 and 20H04727.  S. P. and M. S. were also supported by the MEXT/JSPS KAKENHI No. 15H05888 and No. 15K21733, and by the World Premier International Research Center Initiative (WPI Initiative), MEXT, Japan.


\begin{thebibliography}{99}
\bibitem{Abbott:2016blz} 
B.~P.~Abbott {\it et al.} [LIGO Scientific and Virgo Collaborations],
Phys.\ Rev.\ Lett.\  {\bf 116}, no. 6, 061102 (2016)
doi:10.1103/PhysRevLett.116.061102
[arXiv:1602.03837 [gr-qc]].
\bibitem{Abbott:2016nmj} 
B.~P.~Abbott {\it et al.} [LIGO Scientific and Virgo Collaborations],
Phys.\ Rev.\ Lett.\  {\bf 116}, no. 24, 241103 (2016)
doi:10.1103/PhysRevLett.116.241103
[arXiv:1606.04855 [gr-qc]].
\bibitem{Abbott:2017vtc} 
B.~P.~Abbott {\it et al.} [LIGO Scientific and VIRGO Collaborations],
Phys.\ Rev.\ Lett.\  {\bf 118}, no. 22, 221101 (2017)
doi:10.1103/PhysRevLett.118.221101
[arXiv:1706.01812 [gr-qc]].
\bibitem{Abbott:2017gyy} 
B.~. P.~.Abbott {\it et al.} [LIGO Scientific and Virgo Collaborations],
Astrophys.\ J.\  {\bf 851}, no. 2, L35 (2017)
doi:10.3847/2041-8213/aa9f0c
[arXiv:1711.05578 [astro-ph.HE]].
\bibitem{Abbott:2017oio} 
B.~P.~Abbott {\it et al.} [LIGO Scientific and Virgo Collaborations],
Phys.\ Rev.\ Lett.\  {\bf 119}, no. 14, 141101 (2017)
doi:10.1103/PhysRevLett.119.141101
[arXiv:1709.09660 [gr-qc]].
\bibitem{TheLIGOScientific:2017qsa} 
B.~P.~Abbott {\it et al.} [LIGO Scientific and Virgo Collaborations],
Phys.\ Rev.\ Lett.\  {\bf 119}, no. 16, 161101 (2017)
doi:10.1103/PhysRevLett.119.161101
[arXiv:1710.05832 [gr-qc]].
	

	
	\bibitem{Mandic:2016lcn} 
	V.~Mandic, S.~Bird and I.~Cholis,
	Phys.\ Rev.\ Lett.\  {\bf 117}, no. 20, 201102 (2016)
	doi:10.1103/PhysRevLett.117.201102
	[arXiv:1608.06699 [astro-ph.CO]].
	\bibitem{Clesse:2016ajp} 
	S.~Clesse and J.~García-Bellido,
	Phys.\ Dark Univ.\  {\bf 18}, 105 (2017)
	doi:10.1016/j.dark.2017.10.001
	[arXiv:1610.08479 [astro-ph.CO]].
	\bibitem{Wang:2016ana} 
	S.~Wang, Y.~F.~Wang, Q.~G.~Huang and T.~G.~F.~Li,
	Phys.\ Rev.\ Lett.\  {\bf 120}, no. 19, 191102 (2018)
	doi:10.1103/PhysRevLett.120.191102
	[arXiv:1610.08725 [astro-ph.CO]].
	\bibitem{Raidal:2017mfl} 
	M.~Raidal, V.~Vaskonen and H.~Veerm\"ae,
	JCAP {\bf 1709}, 037 (2017)
	doi:10.1088/1475-7516/2017/09/037
	[arXiv:1707.01480 [astro-ph.CO]].
	\bibitem{Garcia-Bellido:2017aan} 
	~J.~Garcia-Bellido, M.~Peloso and C.~Unal,
	JCAP {\bf 1709}, no. 09, 013 (2017)
	[arXiv:1707.02441 [astro-ph.CO]].
	\bibitem{Guo:2017njn} 
	H.~K.~Guo, J.~Shu and Y.~Zhao,
	Phys.\ Rev.\ D {\bf 99}, no. 2, 023001 (2019)
	doi:10.1103/PhysRevD.99.023001
	[arXiv:1709.03500 [astro-ph.CO]].
	
	
	


	
	
	
	




	


	
	
	
	\bibitem{Witten:1984rs} 
	E.~Witten,
	Phys.\ Rev.\ D {\bf 30}, 272 (1984).
	doi:10.1103/PhysRevD.30.272
	\bibitem{Hogan:1986qda} 
	C.~J.~Hogan,
	Mon.\ Not.\ Roy.\ Astron.\ Soc.\  {\bf 218}, 629 (1986).
	

	\bibitem{Apreda:2001us} 
	R.~Apreda, M.~Maggiore, A.~Nicolis and A.~Riotto,
	Nucl.\ Phys.\ B {\bf 631}, 342 (2002)
	doi:10.1016/S0550-3213(02)00264-X
	[gr-qc/0107033].
	\bibitem{Grojean:2006bp} 
	C.~Grojean and G.~Servant,
	Phys.\ Rev.\ D {\bf 75}, 043507 (2007)
	doi:10.1103/PhysRevD.75.043507
	[hep-ph/0607107].
	\bibitem{Huang:2016odd} 
	F.~P.~Huang, Y.~Wan, D.~G.~Wang, Y.~F.~Cai and X.~Zhang,
	Phys.\ Rev.\ D {\bf 94}, no. 4, 041702 (2016)
	doi:10.1103/PhysRevD.94.041702
	[arXiv:1601.01640 [hep-ph]].
	\bibitem{Cai:2017tmh} 
	R.~G.~Cai, M.~Sasaki and S.~J.~Wang,
	JCAP {\bf 1708}, no. 08, 004 (2017)
	doi:10.1088/1475-7516/2017/08/004
	[arXiv:1707.03001 [astro-ph.CO]].
	\bibitem{Chao:2017vrq} 
	W.~Chao, H.~K.~Guo and J.~Shu,
	JCAP {\bf 1709}, no. 09, 009 (2017)
	doi:10.1088/1475-7516/2017/09/009
	[arXiv:1702.02698 [hep-ph]].
	\bibitem{Wan:2018udw} 
	B.~Imtiaz, Y.~F.~Cai and Y.~Wan,
	Eur.\ Phys.\ J.\ C {\bf 79}, no. 1, 25 (2019)
	doi:10.1140/epjc/s10052-019-6532-y
	[arXiv:1804.05835 [hep-ph]].


\bibitem{Bartolo:2007vp} 
N.~Bartolo, S.~Matarrese, A.~Riotto and A.~Vaihkonen,
Phys.\ Rev.\ D {\bf 76}, 061302 (2007)
doi:10.1103/PhysRevD.76.061302
[arXiv:0705.4240 [astro-ph]].
\bibitem{Biagetti:2013kwa} 
M.~Biagetti, M.~Fasiello and A.~Riotto,
Phys.\ Rev.\ D {\bf 88}, 103518 (2013)
doi:10.1103/PhysRevD.88.103518
[arXiv:1305.7241 [astro-ph.CO]].
\bibitem{Biagetti:2014asa} 
M.~Biagetti, E.~Dimastrogiovanni, M.~Fasiello and M.~Peloso,
JCAP {\bf 1504}, 011 (2015)
doi:10.1088/1475-7516/2015/04/011
[arXiv:1411.3029 [astro-ph.CO]].

\bibitem{Fujita:2014oba} 
T.~Fujita, J.~Yokoyama and S.~Yokoyama,
PTEP {\bf 2015}, 043E01 (2015)
doi:10.1093/ptep/ptv037
[arXiv:1411.3658 [astro-ph.CO]].
\bibitem{Cai:2019jah} 
Y.~F.~Cai, C.~Chen, X.~Tong, D.~G.~Wang and S.~F.~Yan,
arXiv:1902.08187 [astro-ph.CO].
	
	\bibitem{Khlebnikov:1997di} 
	S.~Y.~Khlebnikov and I.~I.~Tkachev,
	Phys.\ Rev.\ D {\bf 56}, 653 (1997)
	doi:10.1103/PhysRevD.56.653
	[hep-ph/9701423].
	\bibitem{GarciaBellido:1998gm} 
	J.~Garcia-Bellido,
	hep-ph/9804205.
	\bibitem{Tashiro:2003qp} 
	H.~Tashiro, T.~Chiba and M.~Sasaki,
	Class.\ Quant.\ Grav.\  {\bf 21}, 1761 (2004)
	doi:10.1088/0264-9381/21/7/004
	[gr-qc/0307068].
	\bibitem{Easther:2006vd} 
	R.~Easther, J.~T.~Giblin, Jr. and E.~A.~Lim,
	Phys.\ Rev.\ Lett.\  {\bf 99}, 221301 (2007)
	doi:10.1103/PhysRevLett.99.221301
	[astro-ph/0612294].
	\bibitem{GarciaBellido:2007dg} 
	J.~Garcia-Bellido and D.~G.~Figueroa,
	Phys.\ Rev.\ Lett.\  {\bf 98}, 061302 (2007)
	doi:10.1103/PhysRevLett.98.061302
	[astro-ph/0701014].
	\bibitem{GarciaBellido:2007af} 
	J.~Garcia-Bellido, D.~G.~Figueroa and A.~Sastre,
	Phys.\ Rev.\ D {\bf 77}, 043517 (2008)
	doi:10.1103/PhysRevD.77.043517
	[arXiv:0707.0839 [hep-ph]].
	\bibitem{Dufaux:2007pt} 
	J.~F.~Dufaux, A.~Bergman, G.~N.~Felder, L.~Kofman and J.~P.~Uzan,
	Phys.\ Rev.\ D {\bf 76}, 123517 (2007)
	doi:10.1103/PhysRevD.76.123517
	[arXiv:0707.0875 [astro-ph]].
	\bibitem{Easther:2007vj} 
	R.~Easther, J.~T.~Giblin and E.~A.~Lim,
	Phys.\ Rev.\ D {\bf 77}, 103519 (2008)
	doi:10.1103/PhysRevD.77.103519
	[arXiv:0712.2991 [astro-ph]].
	\bibitem{Price:2008hq} 
	L.~R.~Price and X.~Siemens,
	Phys.\ Rev.\ D {\bf 78}, 063541 (2008)
	doi:10.1103/PhysRevD.78.063541
	[arXiv:0805.3570 [astro-ph]].
	\bibitem{Jedamzik:2010dq} 
 	K.~Jedamzik, M.~Lemoine and J.~Martin,
 	JCAP {\bf 1009}, 034 (2010)
 	doi:10.1088/1475-7516/2010/09/034
 	[arXiv:1002.3039 [astro-ph.CO]].
	\bibitem{Jedamzik:2010hq} 
	K.~Jedamzik, M.~Lemoine and J.~Martin,
	JCAP {\bf 1004}, 021 (2010)
	doi:10.1088/1475-7516/2010/04/021
	[arXiv:1002.3278 [astro-ph.CO]].
	\bibitem{Kuroyanagi:2015esa} 
	S.~Kuroyanagi, T.~Hiramatsu and J.~Yokoyama,
	JCAP {\bf 1602}, no. 02, 023 (2016)
	doi:10.1088/1475-7516/2016/02/023
	[arXiv:1509.08264 [astro-ph.CO]].
	\bibitem{Liu:2017hua} 
	J.~Liu, Z.~K.~Guo, R.~G.~Cai and G.~Shiu,
	Phys.\ Rev.\ Lett.\  {\bf 120}, no. 3, 031301 (2018)
	doi:10.1103/PhysRevLett.120.031301
	[arXiv:1707.09841 [astro-ph.CO]].
	\bibitem{Kuroyanagi:2017kfx} 
	S.~Kuroyanagi, C.~Lin, M.~Sasaki and S.~Tsujikawa,
	Phys.\ Rev.\ D {\bf 97}, no. 2, 023516 (2018)
	doi:10.1103/PhysRevD.97.023516
	[arXiv:1710.06789 [gr-qc]].
	\bibitem{Amin:2018xfe} 
	M.~A.~Amin, J.~Braden, E.~J.~Copeland, J.~T.~Giblin, C.~Solorio, Z.~J.~Weiner and S.~Y.~Zhou,
	Phys.\ Rev.\ D {\bf 98}, 024040 (2018)
	doi:10.1103/PhysRevD.98.024040
	[arXiv:1803.08047 [astro-ph.CO]].
	\bibitem{Cai:2018tuh} 
	Y.~F.~Cai, X.~Tong, D.~G.~Wang and S.~F.~Yan,
	Phys.\ Rev.\ Lett.\  {\bf 121}, no. 8, 081306 (2018)
	doi:10.1103/PhysRevLett.121.081306
	[arXiv:1805.03639 [astro-ph.CO]].
	\bibitem{Liu:2018rrt} 
	J.~Liu, Z.~K.~Guo, R.~G.~Cai and G.~Shiu,
	Phys.\ Rev.\ D {\bf 99}, no. 10, 103506 (2019)
	doi:10.1103/PhysRevD.99.103506
	[arXiv:1812.09235 [astro-ph.CO]].
	
	


	\bibitem{Binetruy:2012ze} 
	P.~Binetruy, A.~Bohe, C.~Caprini and J.~F.~Dufaux,
	JCAP {\bf 1206}, 027 (2012)
	doi:10.1088/1475-7516/2012/06/027
	[arXiv:1201.0983 [gr-qc]].
	\bibitem{Figueroa:2012kw} 
	D.~G.~Figueroa, M.~Hindmarsh and J.~Urrestilla,
	Phys.\ Rev.\ Lett.\  {\bf 110}, no. 10, 101302 (2013)
	doi:10.1103/PhysRevLett.110.101302
	[arXiv:1212.5458 [astro-ph.CO]].
	
	\bibitem{Gleiser:1998na} 
	M.~Gleiser and R.~Roberts,
	Phys.\ Rev.\ Lett.\  {\bf 81}, 5497 (1998)
	doi:10.1103/PhysRevLett.81.5497
	[astro-ph/9807260].
	\bibitem{Kawasaki:2011vv} 
	M.~Kawasaki and K.~Saikawa,
	JCAP {\bf 1109}, 008 (2011)
	doi:10.1088/1475-7516/2011/09/008
	[arXiv:1102.5628 [astro-ph.CO]].
	\bibitem{Hiramatsu:2013qaa} 
	T.~Hiramatsu, M.~Kawasaki and K.~Saikawa,
	JCAP {\bf 1402}, 031 (2014)
	doi:10.1088/1475-7516/2014/02/031
	[arXiv:1309.5001 [astro-ph.CO]].
	\bibitem{Liu:2019lul} 
	J.~Liu, Z.~K.~Guo and R.~G.~Cai,
	arXiv:1908.02662 [astro-ph.CO].
	

	\bibitem{Durrer:1999bk} 
	R.~Durrer, P.~G.~Ferreira and T.~Kahniashvili,
	Phys.\ Rev.\ D {\bf 61}, 043001 (2000)
	doi:10.1103/PhysRevD.61.043001
	[astro-ph/9911040].
	\bibitem{Caprini:2001nb} 
	C.~Caprini and R.~Durrer,
	Phys.\ Rev.\ D {\bf 65}, 023517 (2001)
	doi:10.1103/PhysRevD.65.023517
	[astro-ph/0106244].
	\bibitem{Pogosian:2001np} 
	L.~Pogosian, T.~Vachaspati and S.~Winitzki,
	Phys.\ Rev.\ D {\bf 65}, 083502 (2002)
	doi:10.1103/PhysRevD.65.083502
	[astro-ph/0112536].
	\bibitem{Caprini:2003vc} 
	C.~Caprini, R.~Durrer and T.~Kahniashvili,
	Phys.\ Rev.\ D {\bf 69}, 063006 (2004)
	doi:10.1103/PhysRevD.69.063006
	[astro-ph/0304556].
	\bibitem{Caprini:2006jb} 
	C.~Caprini and R.~Durrer,
	Phys.\ Rev.\ D {\bf 74}, 063521 (2006)
	doi:10.1103/PhysRevD.74.063521
	[astro-ph/0603476].
	\bibitem{Caprini:2009pr} 
	C.~Caprini, R.~Durrer and E.~Fenu,
	JCAP {\bf 0911}, 001 (2009)
	doi:10.1088/1475-7516/2009/11/001
	[arXiv:0906.4976 [astro-ph.CO]].
	\bibitem{Shaw:2009nf} 
	J.~R.~Shaw and A.~Lewis,
	Phys.\ Rev.\ D {\bf 81}, 043517 (2010)
	doi:10.1103/PhysRevD.81.043517
	[arXiv:0911.2714 [astro-ph.CO]].
	\bibitem{Saga:2018ont} 
	S.~Saga, H.~Tashiro and S.~Yokoyama,
	Phys.\ Rev.\ D {\bf 98}, no. 8, 083518 (2018)
	doi:10.1103/PhysRevD.98.083518
	[arXiv:1807.00561 [astro-ph.CO]].


	
	

	\bibitem{Ade:2015tva}
	P.~A.~R.~Ade {\it et al.} [BICEP2 and Planck Collaborations],
	Phys.\ Rev.\ Lett.\  {\bf 114} (2015) 101301
	doi:10.1103/PhysRevLett.114.101301
	[arXiv:1502.00612 [astro-ph.CO]].
	\bibitem{Li:2017drr} 
	H.~Li {\it et al.},
	Natl.\ Sci.\ Rev.\  {\bf 6}, no. 1, 145 (2019)
	doi:10.1093/nsr/nwy019
	[arXiv:1710.03047 [astro-ph.CO]].
	\bibitem{Matsumura:2013aja} 
	T.~Matsumura {\it et al.},
	J.\ Low.\ Temp.\ Phys.\  {\bf 176}, 733 (2014)
	doi:10.1007/s10909-013-0996-1
	[arXiv:1311.2847 [astro-ph.IM]].


	
\bibitem{Sazhin:1977tq} 
M.~V.~Sazhin,
Vestn.\ Mosk.\ Univ.\ Fiz.\ Astron.\  {\bf 18}, no. 6, 82 (1977).
\bibitem{Detweiler:1979wn} 
S.~L.~Detweiler,
Astrophys.\ J.\  {\bf 234}, 1100 (1979).
doi:10.1086/157593
	
	\bibitem{Desvignes:2016yex} 
	G.~Desvignes {\it et al.},
	Mon.\ Not.\ Roy.\ Astron.\ Soc.\  {\bf 458}, no. 3, 3341 (2016)
	doi:10.1093/mnras/stw483
	[arXiv:1602.08511 [astro-ph.HE]].
	\bibitem{Hobbs:2013aka} 
	G.~Hobbs,
	Class.\ Quant.\ Grav.\  {\bf 30}, 224007 (2013)
	doi:10.1088/0264-9381/30/22/224007
	[arXiv:1307.2629 [astro-ph.IM]].
	\bibitem{McLaughlin:2013ira} 
	M.~A.~McLaughlin,
	Class.\ Quant.\ Grav.\  {\bf 30}, 224008 (2013)
	doi:10.1088/0264-9381/30/22/224008
	[arXiv:1310.0758 [astro-ph.IM]].
	\bibitem{Verbiest:2016vem} 
	J.~P.~W.~Verbiest {\it et al.},
	Mon.\ Not.\ Roy.\ Astron.\ Soc.\  {\bf 458}, no. 2, 1267 (2016)
	doi:10.1093/mnras/stw347
	[arXiv:1602.03640 [astro-ph.IM]].



	\bibitem{Aasi:2013wya} 
	B.~P.~Abbott {\it et al.} [KAGRA and LIGO Scientific and VIRGO Collaborations],
	Living Rev.\ Rel.\  {\bf 21}, no. 1, 3 (2018)
	doi:10.1007/s41114-018-0012-9, 10.1007/lrr-2016-1
	[arXiv:1304.0670 [gr-qc]].
	\bibitem{Punturo:2010zz} 
	M.~Punturo {\it et al.},
	Class.\ Quant.\ Grav.\  {\bf 27}, 194002 (2010).
	doi:10.1088/0264-9381/27/19/194002
	\bibitem{Sathyaprakash:2012jk} 
	B.~Sathyaprakash {\it et al.},
	Class.\ Quant.\ Grav.\  {\bf 29}, 124013 (2012)
	Erratum: [Class.\ Quant.\ Grav.\  {\bf 30}, 079501 (2013)]
	doi:10.1088/0264-9381/29/12/124013, 10.1088/0264-9381/30/7/079501
	[arXiv:1206.0331 [gr-qc]].
	\bibitem{AmaroSeoane:2012km} 
	P.~Amaro-Seoane {\it et al.},
	GW Notes {\bf 6}, 4 (2013)
	[arXiv:1201.3621 [astro-ph.CO]].
	\bibitem{AmaroSeoane:2012je} 
	P.~Amaro-Seoane {\it et al.},
	Class.\ Quant.\ Grav.\  {\bf 29}, 124016 (2012)
	doi:10.1088/0264-9381/29/12/124016
	[arXiv:1202.0839 [gr-qc]].
	\bibitem{Audley:2017drz} 
	H.~Audley {\it et al.} [LISA Collaboration],
	arXiv:1702.00786 [astro-ph.IM].
	\bibitem{Guo:2018npi} 
	Z.~K.~Guo, R.~G.~Cai and Y.~Z.~Zhang,
	arXiv:1807.09495 [gr-qc].
	\bibitem{Luo:2015ght} 
	J.~Luo {\it et al.} [TianQin Collaboration],
	Class.\ Quant.\ Grav.\  {\bf 33}, no. 3, 035010 (2016)
	doi:10.1088/0264-9381/33/3/035010
	[arXiv:1512.02076 [astro-ph.IM]].
	\bibitem{Kawamura:2011zz} 
	S.~Kawamura {\it et al.},
	Class.\ Quant.\ Grav.\  {\bf 28}, 094011 (2011).
	doi:10.1088/0264-9381/28/9/094011
	\bibitem{Crowder:2005nr} 
	J.~Crowder and N.~J.~Cornish,
	Phys.\ Rev.\ D {\bf 72}, 083005 (2005)
	doi:10.1103/PhysRevD.72.083005
	[gr-qc/0506015].
	\bibitem{Corbin:2005ny} 
	V.~Corbin and N.~J.~Cornish,
	Class.\ Quant.\ Grav.\  {\bf 23}, 2435 (2006)
	doi:10.1088/0264-9381/23/7/014
	[gr-qc/0512039].
	\bibitem{Baker:2019pnp} 
	J.~Baker {\it et al.},
	arXiv:1907.11305 [astro-ph.IM].
	
	\bibitem{Kuroyanagi:2018csn} 
	S.~Kuroyanagi, T.~Chiba and T.~Takahashi,
	JCAP {\bf 1811}, no. 11, 038 (2018)
	doi:10.1088/1475-7516/2018/11/038
	[arXiv:1807.00786 [astro-ph.CO]].
	\bibitem{Caprini:2019pxz} 
	C.~Caprini, D.~G.~Figueroa, R.~Flauger, G.~Nardini, M.~Peloso, M.~Pieroni, A.~Ricciardone and G.~Tasinato,
	arXiv:1906.09244 [astro-ph.CO].


	\bibitem{Phinney:2001di} 
	E.~S.~Phinney,
	astro-ph/0108028.
	\bibitem{Regimbau:2011rp} 
	T.~Regimbau,
	Res.\ Astron.\ Astrophys.\  {\bf 11}, 369 (2011)
	doi:10.1088/1674-4527/11/4/001
	[arXiv:1101.2762 [astro-ph.CO]].
	\bibitem{Zhu:2011bd} 
	X.~J.~Zhu, E.~Howell, T.~Regimbau, D.~Blair and Z.~H.~Zhu,
	Astrophys.\ J.\  {\bf 739}, 86 (2011)
	doi:10.1088/0004-637X/739/2/86
	[arXiv:1104.3565 [gr-qc]].
	\bibitem{Rosado:2011kv} 
	P.~A.~Rosado,
	Phys.\ Rev.\ D {\bf 84}, 084004 (2011)
	doi:10.1103/PhysRevD.84.084004
	[arXiv:1106.5795 [gr-qc]].
	\bibitem{Marassi:2011si} 
	S.~Marassi, R.~Schneider, G.~Corvino, V.~Ferrari and S.~Portegies Zwart,
	Phys.\ Rev.\ D {\bf 84}, 124037 (2011)
	doi:10.1103/PhysRevD.84.124037
	[arXiv:1111.6125 [astro-ph.CO]].
	\bibitem{Zhu:2012xw} 
	X.~J.~Zhu, E.~J.~Howell, D.~G.~Blair and Z.~H.~Zhu,
	Mon.\ Not.\ Roy.\ Astron.\ Soc.\  {\bf 431}, no. 1, 882 (2013)
	doi:10.1093/mnras/stt207
	[arXiv:1209.0595 [gr-qc]].

	\bibitem{Caprini:2009fx} 
	C.~Caprini, R.~Durrer, T.~Konstandin and G.~Servant,
	Phys.\ Rev.\ D {\bf 79}, 083519 (2009)
	doi:10.1103/PhysRevD.79.083519
	[arXiv:0901.1661 [astro-ph.CO]].
	\bibitem{Hindmarsh:2016lnk} 
	M.~Hindmarsh,
	Phys.\ Rev.\ Lett.\  {\bf 120}, no. 7, 071301 (2018)
	doi:10.1103/PhysRevLett.120.071301
	[arXiv:1608.04735 [astro-ph.CO]].
	\bibitem{Hindmarsh:2017gnf} 
	M.~Hindmarsh, S.~J.~Huber, K.~Rummukainen and D.~J.~Weir,
	Phys.\ Rev.\ D {\bf 96}, no. 10, 103520 (2017)
	doi:10.1103/PhysRevD.96.103520
	[arXiv:1704.05871 [astro-ph.CO]].
	\bibitem{Inomata:2019zqy} 
	K.~Inomata, K.~Kohri, T.~Nakama and T.~Terada,
	arXiv:1904.12878 [astro-ph.CO].
	\bibitem{Inomata:2019ivs} 
	K.~Inomata, K.~Kohri, T.~Nakama and T.~Terada,
	Phys.\ Rev.\ D {\bf 100}, no. 4, 043532 (2019)
	doi:10.1103/PhysRevD.100.043532
	[arXiv:1904.12879 [astro-ph.CO]].
	\bibitem{Hindmarsh:2019phv} 
	M.~Hindmarsh and M.~Hijazi,
	arXiv:1909.10040 [astro-ph.CO].












	

	
	
	
	
	
	
	
	\bibitem{Ananda:2006af} 
	K.~N.~Ananda, C.~Clarkson and D.~Wands,
	Phys.\ Rev.\ D {\bf 75}, 123518 (2007)
	doi:10.1103/PhysRevD.75.123518
	[gr-qc/0612013].
	\bibitem{Baumann:2007zm} 
	D.~Baumann, P.~J.~Steinhardt, K.~Takahashi and K.~Ichiki,
	Phys.\ Rev.\ D {\bf 76}, 084019 (2007)
	doi:10.1103/PhysRevD.76.084019
	[hep-th/0703290].
	\bibitem{Saito:2008jc} 
	R.~Saito and J.~Yokoyama,
	Phys.\ Rev.\ Lett.\  {\bf 102}, 161101 (2009)
	Erratum: [Phys.\ Rev.\ Lett.\  {\bf 107}, 069901 (2011)]
	doi:10.1103/PhysRevLett.102.161101, 10.1103/PhysRevLett.107.069901
	[arXiv:0812.4339 [astro-ph]].
	\bibitem{Assadullahi:2009nf} 
	H.~Assadullahi and D.~Wands,
	Phys.\ Rev.\ D {\bf 79}, 083511 (2009)
	doi:10.1103/PhysRevD.79.083511
	[arXiv:0901.0989 [astro-ph.CO]].
	\bibitem{Assadullahi:2009jc} 
	H.~Assadullahi and D.~Wands,
	Phys.\ Rev.\ D {\bf 81}, 023527 (2010)
	doi:10.1103/PhysRevD.81.023527
	[arXiv:0907.4073 [astro-ph.CO]].
	\bibitem{Bugaev:2009zh} 
	E.~Bugaev and P.~Klimai,
	Phys.\ Rev.\ D {\bf 81}, 023517 (2010)
	doi:10.1103/PhysRevD.81.023517
	[arXiv:0908.0664 [astro-ph.CO]].
	\bibitem{Saito:2009jt} 
	R.~Saito and J.~Yokoyama,
	Prog.\ Theor.\ Phys.\  {\bf 123}, 867 (2010)
	Erratum: [Prog.\ Theor.\ Phys.\  {\bf 126}, 351 (2011)]
	doi:10.1143/PTP.126.351, 10.1143/PTP.123.867
	[arXiv:0912.5317 [astro-ph.CO]].
	\bibitem{Bugaev:2010bb} 
	E.~Bugaev and P.~Klimai,
	Phys.\ Rev.\ D {\bf 83}, 083521 (2011)
	doi:10.1103/PhysRevD.83.083521
	[arXiv:1012.4697 [astro-ph.CO]].
	\bibitem{Alabidi:2012ex} 
	L.~Alabidi, K.~Kohri, M.~Sasaki and Y.~Sendouda,
	JCAP {\bf 1209}, 017 (2012)
	doi:10.1088/1475-7516/2012/09/017
	[arXiv:1203.4663 [astro-ph.CO]].
	\bibitem{Alabidi:2013wtp} 
	~L.~Alabidi, K.~Kohri, M.~Sasaki and Y.~Sendouda,
	JCAP {\bf 1305}, 033 (2013)
	[arXiv:1303.4519 [astro-ph.CO]].
	\bibitem{Inomata:2016rbd} 
	K.~Inomata, M.~Kawasaki, K.~Mukaida, Y.~Tada and T.~T.~Yanagida,
	Phys.\ Rev.\ D {\bf 95}, no. 12, 123510 (2017)
	doi:10.1103/PhysRevD.95.123510
	[arXiv:1611.06130 [astro-ph.CO]].
	\bibitem{Orlofsky:2016vbd} 
	N.~Orlofsky, A.~Pierce and J.~D.~Wells,
	Phys.\ Rev.\ D {\bf 95}, no. 6, 063518 (2017)
	doi:10.1103/PhysRevD.95.063518
	[arXiv:1612.05279 [astro-ph.CO]].
	\bibitem{Nakama:2016gzw}
	T.~Nakama, J.~Silk and M.~Kamionkowski,
	Phys.\ Rev.\ D {\bf 95} (2017) no.4,  043511
	doi:10.1103/PhysRevD.95.043511
	[arXiv:1612.06264 [astro-ph.CO]].
	\bibitem{Gong:2017qlj} 
	H. Di and Y.~Gong,
	JCAP {\bf 1807}, no. 07, 007 (2018)
	doi:10.1088/1475-7516/2018/07/007
	[arXiv:1707.09578 [astro-ph.CO]].
	\bibitem{Espinosa:2018eve}
	J.~R.~Espinosa, D.~Racco and A.~Riotto,
	JCAP {\bf 1809} (2018) no.09,  012
	doi:10.1088/1475-7516/2018/09/012
	[arXiv:1804.07732 [hep-ph]].
	\bibitem{Kohri:2018awv} 
	K.~Kohri and T.~Terada,
	Phys.\ Rev.\ D {\bf 97}, 123532 (2018)
	doi:10.1103/PhysRevD.97.123532
	[arXiv:1804.08577 [gr-qc]].
	
	
	
	
	\bibitem{Cai:2018dig} 
	R.~g.~Cai, S.~Pi and M.~Sasaki,
	Phys.\ Rev.\ Lett.\  {\bf 122}, no. 20, 201101 (2019)
	doi:10.1103/PhysRevLett.122.201101
	[arXiv:1810.11000 [astro-ph.CO]].
	\bibitem{Bartolo:2018evs} 
	N.~Bartolo, V.~De Luca, G.~Franciolini, A.~Lewis, M.~Peloso and A.~Riotto,
	Phys.\ Rev.\ Lett.\  {\bf 122}, no. 21, 211301 (2019)
	doi:10.1103/PhysRevLett.122.211301
	[arXiv:1810.12218 [astro-ph.CO]].
	\bibitem{Bartolo:2018rku} 
	N.~Bartolo, V.~De Luca, G.~Franciolini, M.~Peloso, D.~Racco and A.~Riotto,
	Phys.\ Rev.\ D {\bf 99}, no. 10, 103521 (2019)
	doi:10.1103/PhysRevD.99.103521
	[arXiv:1810.12224 [astro-ph.CO]].
	\bibitem{Unal:2018yaa} 
	C.~Unal,
	Phys.\ Rev.\ D {\bf 99}, no. 4, 041301 (2019)
	doi:10.1103/PhysRevD.99.041301
	[arXiv:1811.09151 [astro-ph.CO]].
	\bibitem{Byrnes:2018txb} 
	C.~T.~Byrnes, P.~S.~Cole and S.~P.~Patil,
	arXiv:1811.11158 [astro-ph.CO].
	\bibitem{Inomata:2018epa} 
	K.~Inomata and T.~Nakama,
	Phys.\ Rev.\ D {\bf 99}, no. 4, 043511 (2019)
	doi:10.1103/PhysRevD.99.043511
	[arXiv:1812.00674 [astro-ph.CO]].
	\bibitem{Dalianis:2018ymb} 
	I.~Dalianis,
	arXiv:1812.09807 [astro-ph.CO].
	\bibitem{Cai:2019amo} 
	R.~G.~Cai, S.~Pi, S.~J.~Wang and X.~Y.~Yang,
	JCAP {\bf 1905}, no. 05, 013 (2019)
	doi:10.1088/1475-7516/2019/05/013
	[arXiv:1901.10152 [astro-ph.CO]].
	\bibitem{Wang:2019kaf} 
	S.~Wang, T.~Terada and K.~Kohri,
	Phys.\ Rev.\ D {\bf 99}, no. 10, 103531 (2019)
	doi:10.1103/PhysRevD.99.103531
	[arXiv:1903.05924 [astro-ph.CO]].
	\bibitem{DeLuca:2019qsy} 
	V.~De Luca, G.~Franciolini, A.~Kehagias, M.~Peloso, A.~Riotto and C.~Ünal,
	arXiv:1904.00970 [astro-ph.CO].
	\bibitem{Tada:2019amh} 
	Y.~Tada and S.~Yokoyama,
	arXiv:1904.10298 [astro-ph.CO].
	\bibitem{Yuan:2019udt} 
	C.~Yuan, Z.~C.~Chen and Q.~G.~Huang,
	arXiv:1906.11549 [astro-ph.CO].
	\bibitem{Xu:2019bdp} 
	W.~T.~Xu, J.~Liu, T.~J.~Gao and Z.~K.~Guo,
	arXiv:1907.05213 [astro-ph.CO].
	\bibitem{Cai:2019elf} 
	R.~G.~Cai, S.~Pi, S.~J.~Wang and X.~Y.~Yang,
	arXiv:1907.06372 [astro-ph.CO].
	\bibitem{Lu:2019sti} 
	Y.~Lu, Y.~Gong, Z.~Yi and F.~Zhang,
	arXiv:1907.11896 [gr-qc].
	\bibitem{Kalaja:2019uju} 
	A.~Kalaja, N.~Bellomo, N.~Bartolo, D.~Bertacca, S.~Matarrese, I.~Musco, A.~Raccanelli and L.~Verde,
	arXiv:1908.03596 [astro-ph.CO].
	\bibitem{Chen:2019zza} 
	C.~Chen and Y.~F.~Cai,
	arXiv:1908.03942 [astro-ph.CO].
	\bibitem{Atal:2019erb} 
	V.~Atal, J.~Cid, A.~Escrivà and J.~Garriga,
	arXiv:1908.11357 [astro-ph.CO].
	\bibitem{Gong:2019mui} 
	J.~O.~Gong,
	arXiv:1909.12708 [gr-qc].
	
\bibitem{Akrami:2018odb} 
Y.~Akrami {\it et al.} [Planck Collaboration],
arXiv:1807.06211 [astro-ph.CO].



	
	
	\bibitem{Zeldovich:1963}
	Ya. B. Zel'dovich and I.D. Novikov, Sov. Astron. {\bf{10}}, 602 (1966).
	\bibitem{Hawking:1971ei} 
	S.~Hawking,
	Mon.\ Not.\ Roy.\ Astron.\ Soc.\  {\bf 152}, 75 (1971).
	\bibitem{Carr:1974nx} 
	B.~J.~Carr and S.~W.~Hawking,
	Mon.\ Not.\ Roy.\ Astron.\ Soc.\  {\bf 168}, 399 (1974).
	\bibitem{Meszaros:1974tb} 
	P.~Meszaros,
	Astron.\ Astrophys.\  {\bf 37}, 225 (1974).
	\bibitem{Carr:1975qj} 
	B.~J.~Carr,
	Astrophys.\ J.\  {\bf 201}, 1 (1975).
	

	\bibitem{Sasaki:2018dmp} 
	M.~Sasaki, T.~Suyama, T.~Tanaka and S.~Yokoyama,
	Class.\ Quant.\ Grav.\  {\bf 35}, no. 6, 063001 (2018)
	doi:10.1088/1361-6382/aaa7b4
	[arXiv:1801.05235 [astro-ph.CO]].
	
	
	\bibitem{Green:2004wb} 
	A.~M.~Green, A.~R.~Liddle, K.~A.~Malik and M.~Sasaki,
	Phys.\ Rev.\ D {\bf 70}, 041502 (2004)
	[astro-ph/0403181].
	\bibitem{Frampton:2009nx} 
	P.~H.~Frampton,
	JCAP {\bf 0910}, 016 (2009)
	[arXiv:0905.3632 [hep-th]].
	\bibitem{Carr:2009jm} 
	~B.~J.~Carr, K.~Kohri, Y.~Sendouda and J.~Yokoyama,
	Phys.\ Rev.\ D {\bf 81}, 104019 (2010)
	[arXiv:0912.5297 [astro-ph.CO]].
	\bibitem{Carr:2016hva} 
	B.~J.~Carr, K.~Kohri, Y.~Sendouda and J.~Yokoyama,
	Phys.\ Rev.\ D {\bf 94}, no. 4, 044029 (2016)
	doi:10.1103/PhysRevD.94.044029
	[arXiv:1604.05349 [astro-ph.CO]].
	\bibitem{Carr:2016drx} 
	B.~Carr, F.~Kuhnel and M.~Sandstad,
	Phys.\ Rev.\ D {\bf 94}, no. 8, 083504 (2016)
	[arXiv:1607.06077 [astro-ph.CO]].
	\bibitem{Poulter:2019ooo} 
	H.~Poulter, Y.~Ali-Haïmoud, J.~Hamann, M.~White and A.~G.~Williams,
	arXiv:1907.06485 [astro-ph.CO].

	\bibitem{Young:2014ana} 
	S.~Young, C.~T.~Byrnes and M.~Sasaki,
	JCAP {\bf 1407}, 045 (2014)
	doi:10.1088/1475-7516/2014/07/045
	[arXiv:1405.7023 [gr-qc]].
	\bibitem{Byrnes:2018clq} 
	C.~T.~Byrnes, M.~Hindmarsh, S.~Young and M.~R.~S.~Hawkins,
	JCAP {\bf 1808}, no. 08, 041 (2018)
	doi:10.1088/1475-7516/2018/08/041
	[arXiv:1801.06138 [astro-ph.CO]].


	\bibitem{Niikura:2017zjd} 
	H.~Niikura {\it et al.},
	arXiv:1701.02151 [astro-ph.CO].
	\bibitem{Katz:2018zrn} 
	A.~Katz, J.~Kopp, S.~Sibiryakov and W.~Xue,
	[arXiv:1807.11495 [astro-ph.CO]].
	\bibitem{Montero-Camacho:2019jte} 
	P.~Montero-Camacho, X.~Fang, G.~Vasquez, M.~Silva and C.~M.~Hirata,
	arXiv:1906.05950 [astro-ph.CO].
	

	\bibitem{Tisserand:2006zx} 
	P.~Tisserand {\it et al.} [EROS-2 Collaboration],
	Astron.\ Astrophys.\  {\bf 469}, 387 (2007)
	doi:10.1051/0004-6361:20066017
	[astro-ph/0607207].
	\bibitem{Graham:2015apa} 
	P.~W.~Graham, S.~Rajendran and J.~Varela,
	Phys.\ Rev.\ D {\bf 92}, no. 6, 063007 (2015)
	doi:10.1103/PhysRevD.92.063007
	[arXiv:1505.04444 [hep-ph]].
	\bibitem{Koushiappas:2017chw} 
	S.~M.~Koushiappas and A.~Loeb,
	Phys.\ Rev.\ Lett.\  {\bf 119}, no. 4, 041102 (2017)
	doi:10.1103/PhysRevLett.119.041102
	[arXiv:1704.01668 [astro-ph.GA]].
	\bibitem{Authors:2019qbw} 
	B.~P.~Abbott {\it et al.} [LIGO Scientific and Virgo Collaborations],
	arXiv:1904.08976 [astro-ph.CO].
	
	
	\bibitem{Frampton:2010sw} 
	P.~H.~Frampton, M.~Kawasaki, F.~Takahashi and T.~T.~Yanagida,
	JCAP {\bf 1004}, 023 (2010)
	doi:10.1088/1475-7516/2010/04/023
	[arXiv:1001.2308 [hep-ph]].
	\bibitem{Kawasaki:2012wr} 
	M.~Kawasaki, N.~Kitajima and T.~T.~Yanagida,
	Phys.\ Rev.\ D {\bf 87}, no. 6, 063519 (2013)
	doi:10.1103/PhysRevD.87.063519
	[arXiv:1207.2550 [hep-ph]].
	\bibitem{Pi:2017gih} 
	S.~Pi, Y.~l.~Zhang, Q.~G.~Huang and M.~Sasaki,
	JCAP {\bf 1805}, no. 05, 042 (2018)
	doi:10.1088/1475-7516/2018/05/042
	[arXiv:1712.09896 [astro-ph.CO]].
	
	
	
	
	
	\bibitem{GarciaBellido:1996qt} 
	J.~Garcia-Bellido, A.~D.~Linde and D.~Wands,
	Phys.\ Rev.\ D {\bf 54}, 6040 (1996)
	doi:10.1103/PhysRevD.54.6040
	[astro-ph/9605094].
	\bibitem{Kawasaki:1997ju} 
	M.~Kawasaki, N.~Sugiyama and T.~Yanagida,
	Phys.\ Rev.\ D {\bf 57}, 6050 (1998)
	doi:10.1103/PhysRevD.57.6050
	[hep-ph/9710259].
	\bibitem{Yokoyama:1998pt} 
	J.~Yokoyama,
	Phys.\ Rev.\ D {\bf 58}, 083510 (1998)
	doi:10.1103/PhysRevD.58.083510
	[astro-ph/9802357].
	\bibitem{Kohri:2012yw} 
	K.~Kohri, C.~M.~Lin and T.~Matsuda,
	Phys.\ Rev.\ D {\bf 87}, no. 10, 103527 (2013)
	doi:10.1103/PhysRevD.87.103527
	[arXiv:1211.2371 [hep-ph]].
	\bibitem{Clesse:2015wea} 
	S.~Clesse and J.~Garcia-Bellido,
	Phys.\ Rev.\ D {\bf 92}, no. 2, 023524 (2015)
	doi:10.1103/PhysRevD.92.023524
	[arXiv:1501.07565 [astro-ph.CO]].
	\bibitem{Cheng:2016qzb}
	S.~L.~Cheng, W.~Lee and K.~W.~Ng,
	JHEP {\bf 1702} (2017) 008
	doi:10.1007/JHEP02(2017)008
	[arXiv:1606.00206 [astro-ph.CO]].
	\bibitem{Inomata:2017okj} 
	K.~Inomata, M.~Kawasaki, K.~Mukaida, Y.~Tada and T.~T.~Yanagida,
	Phys.\ Rev.\ D {\bf 96}, no. 4, 043504 (2017)
	doi:10.1103/PhysRevD.96.043504
	[arXiv:1701.02544 [astro-ph.CO]].
	\bibitem{Garcia-Bellido:2017mdw} 
	J.~Garcia-Bellido and E.~Ruiz Morales,
	Phys.\ Dark Univ.\  {\bf 18}, 47 (2017)
	doi:10.1016/j.dark.2017.09.007
	[arXiv:1702.03901 [astro-ph.CO]].
	\bibitem{Kannike:2017bxn} 
	K.~Kannike, L.~Marzola, M.~Raidal and H.~Veerm\"{a}e,
	JCAP {\bf 1709}, no. 09, 020 (2017)
	doi:10.1088/1475-7516/2017/09/020
	[arXiv:1705.06225 [astro-ph.CO]].
	\bibitem{Inomata:2017vxo} 
	K.~Inomata, M.~Kawasaki, K.~Mukaida and T.~T.~Yanagida,
	Phys.\ Rev.\ D {\bf 97}, no. 4, 043514 (2018)
	doi:10.1103/PhysRevD.97.043514
	[arXiv:1711.06129 [astro-ph.CO]].
	\bibitem{Ando:2017veq} 
	K.~Ando, K.~Inomata, M.~Kawasaki, K.~Mukaida and T.~T.~Yanagida,
	Phys.\ Rev.\ D {\bf 97}, no. 12, 123512 (2018)
	doi:10.1103/PhysRevD.97.123512
	[arXiv:1711.08956 [astro-ph.CO]].
	\bibitem{Cheng:2018yyr} 
	S.~L.~Cheng, W.~Lee and K.~W.~Ng,
	JCAP {\bf 1807}, no. 07, 001 (2018)
	doi:10.1088/1475-7516/2018/07/001
	[arXiv:1801.09050 [astro-ph.CO]].
	\bibitem{Ando:2018nge} 
	K.~Ando, M.~Kawasaki and H.~Nakatsuka,
	[arXiv:1805.07757 [astro-ph.CO]].
	\bibitem{Espinosa:2017sgp} 
	J.~R.~Espinosa, D.~Racco and A.~Riotto,
	Phys.\ Rev.\ Lett.\  {\bf 120}, no. 12, 121301 (2018),~doi:10.1103/PhysRevLett.120.121301
	[arXiv:1710.11196 [hep-ph]].
	
	
	\bibitem{Bird:2016dcv} 
	S.~Bird, I.~Cholis, J.~B.~Mu\~noz, Y.~Ali-Ha\"imoud, M.~Kamionkowski, E.~D.~Kovetz, A.~Raccanelli and A.~G.~Riess,
	Phys.\ Rev.\ Lett.\  {\bf 116}, no. 20, 201301 (2016)
	[arXiv:1603.00464 [astro-ph.CO]].
	\bibitem{Clesse:2016vqa} 
	~S.~Clesse and J.~Garcia-Bellido,
	Phys.\ Dark Univ.\  {\bf 15}, 142 (2017)
	[arXiv:1603.05234 [astro-ph.CO]].
	\bibitem{Sasaki:2016jop} 
	~M.~Sasaki, T.~Suyama, T.~Tanaka and S.~Yokoyama,
	Phys.\ Rev.\ Lett.\  {\bf 117}, no. 6, 061101 (2016)
	[arXiv:1603.08338 [astro-ph.CO]].
	\bibitem{Chen:2016pud} 
	~L.~Chen, Q.~G.~Huang and K.~Wang,
	JCAP {\bf 1612}, no. 12, 044 (2016)
	[arXiv:1608.02174 [astro-ph.CO]].
	\bibitem{Blinnikov:2016bxu} 
	S.~Blinnikov, A.~Dolgov, N.~K.~Porayko and K.~Postnov,
	JCAP {\bf 1611}, no. 11, 036 (2016)
	[arXiv:1611.00541 [astro-ph.HE]].
	\bibitem{Ali-Haimoud:2016mbv} 
	~Y.~Ali-Haimoud and M.~Kamionkowski,
	Phys.\ Rev.\ D {\bf 95}, no. 4, 043534 (2017)
	[arXiv:1612.05644 [astro-ph.CO]].
	\bibitem{Zumalacarregui:2017qqd} 
	M.~Zumalacarregui and U.~Seljak,
	Phys.\ Rev.\ Lett.\  {\bf 121}, no. 14, 141101 (2018)
	doi:10.1103/PhysRevLett.121.141101
	[arXiv:1712.02240 [astro-ph.CO]].
	\bibitem{Garcia-Bellido:2017imq} 
	J.~Garcia-Bellido, S.~Clesse and P.~Fleury,
	Phys.\ Dark Univ.\  {\bf 20}, 95 (2018)
	doi:10.1016/j.dark.2018.04.005
	[arXiv:1712.06574 [astro-ph.CO]].
	
	
	\bibitem{Bean:2002kx} 
	R.~Bean and J.~Magueijo,
	Phys.\ Rev.\ D {\bf 66}, 063505 (2002)
	doi:10.1103/PhysRevD.66.063505
	[astro-ph/0204486].
	\bibitem{Kawasaki:2012kn} 
	M.~Kawasaki, A.~Kusenko and T.~T.~Yanagida,
	Phys.\ Lett.\ B {\bf 711}, 1 (2012)
	doi:10.1016/j.physletb.2012.03.056
	[arXiv:1202.3848 [astro-ph.CO]].
	\bibitem{Nakama:2017xvq} 
	T.~Nakama, B.~Carr and J.~Silk,
	Phys.\ Rev.\ D {\bf 97}, no. 4, 043525 (2018)
	doi:10.1103/PhysRevD.97.043525
	[arXiv:1710.06945 [astro-ph.CO]].
	\bibitem{Carr:2018rid} 
	B.~Carr and J.~Silk,
	Mon.\ Not.\ Roy.\ Astron.\ Soc.\  {\bf 478}, no. 3, 3756 (2018)
	doi:10.1093/mnras/sty1204
	[arXiv:1801.00672 [astro-ph.CO]].
	\bibitem{Nakama:2019htb} 
	T.~Nakama, K.~Kohri and J.~Silk,
	Phys.\ Rev.\ D {\bf 99}, no. 12, 123530 (2019)
	doi:10.1103/PhysRevD.99.123530
	[arXiv:1905.04477 [astro-ph.CO]].
	
	
	
	
	
	
	
	
	
	
	
	
	

%
%
%
%
	
	
	
	
	
	
	
	
	
	
	
%
%
%
%
%
%
%
%
%
%
%
%
%
%
%
%
%
%
%
%
%
%
%
%
%
%
%
%
%

	
	
	
	
	
	
	
	
%
%
%
%
%
%
	
	


\bibitem{Turner:1990rc} 
M.~S.~Turner and F.~Wilczek,
Phys.\ Rev.\ Lett.\  {\bf 65}, 3080 (1990).
doi:10.1103/PhysRevLett.65.3080
\bibitem{Kosowsky:1991ua} 
A.~Kosowsky, M.~S.~Turner and R.~Watkins,
Phys.\ Rev.\ D {\bf 45}, 4514 (1992).
doi:10.1103/PhysRevD.45.4514
\bibitem{Kosowsky:1992rz} 
A.~Kosowsky, M.~S.~Turner and R.~Watkins,
Phys.\ Rev.\ Lett.\  {\bf 69}, 2026 (1992).
doi:10.1103/PhysRevLett.69.2026
\bibitem{Kosowsky:1992vn} 
A.~Kosowsky and M.~S.~Turner,
Phys.\ Rev.\ D {\bf 47}, 4372 (1993)
doi:10.1103/PhysRevD.47.4372
[astro-ph/9211004].
\bibitem{Turner:1992tz} 
M.~S.~Turner, E.~J.~Weinberg and L.~M.~Widrow,
Phys.\ Rev.\ D {\bf 46}, 2384 (1992).
doi:10.1103/PhysRevD.46.2384
\bibitem{Kamionkowski:1993fg} 
M.~Kamionkowski, A.~Kosowsky and M.~S.~Turner,
Phys.\ Rev.\ D {\bf 49}, 2837 (1994)
doi:10.1103/PhysRevD.49.2837
[astro-ph/9310044].
\bibitem{Caprini:2007xq} 
C.~Caprini, R.~Durrer and G.~Servant,
Phys.\ Rev.\ D {\bf 77}, 124015 (2008)
doi:10.1103/PhysRevD.77.124015
[arXiv:0711.2593 [astro-ph]].
\bibitem{Jinno:2016vai} 
R.~Jinno and M.~Takimoto,
Phys.\ Rev.\ D {\bf 95}, no. 2, 024009 (2017)
doi:10.1103/PhysRevD.95.024009
[arXiv:1605.01403 [astro-ph.CO]].
\bibitem{Cutting:2018tjt} 
D.~Cutting, M.~Hindmarsh and D.~J.~Weir,
Phys.\ Rev.\ D {\bf 97}, no. 12, 123513 (2018)
doi:10.1103/PhysRevD.97.123513
[arXiv:1802.05712 [astro-ph.CO]].

\bibitem{Hindmarsh:2013xza} 
M.~Hindmarsh, S.~J.~Huber, K.~Rummukainen and D.~J.~Weir,
Phys.\ Rev.\ Lett.\  {\bf 112}, 041301 (2014)
doi:10.1103/PhysRevLett.112.041301
[arXiv:1304.2433 [hep-ph]].
\bibitem{Hindmarsh:2015qta} 
M.~Hindmarsh, S.~J.~Huber, K.~Rummukainen and D.~J.~Weir,
Phys.\ Rev.\ D {\bf 92}, no. 12, 123009 (2015)
doi:10.1103/PhysRevD.92.123009
[arXiv:1504.03291 [astro-ph.CO]].

\bibitem{Kosowsky:2001xp} 
A.~Kosowsky, A.~Mack and T.~Kahniashvili,
Phys.\ Rev.\ D {\bf 66}, 024030 (2002)
doi:10.1103/PhysRevD.66.024030
[astro-ph/0111483].
\bibitem{Dolgov:2002ra} 
A.~D.~Dolgov, D.~Grasso and A.~Nicolis,
Phys.\ Rev.\ D {\bf 66}, 103505 (2002)
doi:10.1103/PhysRevD.66.103505
[astro-ph/0206461].
\bibitem{Kahniashvili:2005qi} 
T.~Kahniashvili, G.~Gogoberidze and B.~Ratra,
Phys.\ Rev.\ Lett.\  {\bf 95}, 151301 (2005)
doi:10.1103/PhysRevLett.95.151301
[astro-ph/0505628].
\bibitem{Gogoberidze:2007an} 
G.~Gogoberidze, T.~Kahniashvili and A.~Kosowsky,
Phys.\ Rev.\ D {\bf 76}, 083002 (2007)
doi:10.1103/PhysRevD.76.083002
[arXiv:0705.1733 [astro-ph]].
\bibitem{Kahniashvili:2008pe} 
T.~Kahniashvili, L.~Campanelli, G.~Gogoberidze, Y.~Maravin and B.~Ratra,
Phys.\ Rev.\ D {\bf 78}, 123006 (2008)
Erratum: [Phys.\ Rev.\ D {\bf 79}, 109901 (2009)]
doi:10.1103/PhysRevD.78.123006, 10.1103/PhysRevD.79.109901
[arXiv:0809.1899 [astro-ph]].
\bibitem{Caprini:2009yp} 
C.~Caprini, R.~Durrer and G.~Servant,
JCAP {\bf 0912}, 024 (2009)
doi:10.1088/1475-7516/2009/12/024
[arXiv:0909.0622 [astro-ph.CO]].


	

\bibitem{Nakamura:1997sm} 
T.~Nakamura, M.~Sasaki, T.~Tanaka and K.~S.~Thorne,
Astrophys.\ J.\  {\bf 487}, L139 (1997)
doi:10.1086/310886
[astro-ph/9708060].



	
	
	\bibitem{Linde:1996gt} 
	A.~D.~Linde and V.~F.~Mukhanov,
	Phys.\ Rev.\ D {\bf 56}, R535 (1997)
	doi:10.1103/PhysRevD.56.R535
	[astro-ph/9610219].
	\bibitem{Liddle:1999hq} 
	A.~R.~Liddle, D.~H.~Lyth, K.~A.~Malik and D.~Wands,
	Phys.\ Rev.\ D {\bf 61}, 103509 (2000)
	doi:10.1103/PhysRevD.61.103509
	[hep-ph/9912473].
	\bibitem{Suyama:2004mz} 
	T.~Suyama, T.~Tanaka, B.~Bassett and H.~Kudoh,
	Phys.\ Rev.\ D {\bf 71}, 063507 (2005)
	doi:10.1103/PhysRevD.71.063507
	[hep-ph/0410247].
\end{thebibliography}
 \end{document}